\documentclass{PoS}

\usepackage{bm}
\usepackage[super,sort&compress]{natbib}
\usepackage{amsmath}
\usepackage{amssymb}
\usepackage{amsfonts}

\newcommand{\ie}{\emph{i.e.} }

\newcommand{\fin}{\mbox{ .}}

\newcommand{\mytautheta}{\tau}
\newcommand{\mytaur}{\varrho}

\newcommand{\cm}{\mbox{ cm}}
\newcommand{\sr}{\mbox{ sr}}
\newcommand{\se}{\mbox{ s}}
\newcommand{\yr}{\mbox{ yr}}

\newcommand{\erg}{\mbox{ erg}}

\newcommand{\keV}{\mbox{ keV}}

\newcommand{\GeV}{\mbox{ GeV}}
\newcommand{\TeV}{\mbox{ TeV}}

\newcommand{\K}{\mbox{ K}}
\newcommand{\dgr}{^{\circ}}

\newcommand{\const}{\mbox{const}}

\newcommand{\gama}{$\gamma$}

\newcommand{\myNi}{\emph{(i)}\,}
\newcommand{\myNii}{\emph{(ii)}\,}
\newcommand{\myNiii}{\emph{(iii)}\,}
\newcommand{\myNiv}{\emph{(iv)}\,}

\newcommand{\till}{{\mbox{--}}}

\newcommand{\mySagn}{s_p}
\newcommand{\myLAGN}{L_p}

\newcommand{\myR}{R_{500}}

\newcommand{\mynewcommand}[2]{\ifdefined #1 \else \newcommand{#1}{#2} \fi}
\mynewcommand\pasj{{PASJ}}   
\mynewcommand{\apj}{ApJ}     
\mynewcommand{\apjl}{ApJL}     
\mynewcommand{\apjs}{ApJS}    
\mynewcommand\mnras{{MNRAS}} 
\mynewcommand{\aap}{A\&A}    
\mynewcommand{\nat}{Nature}  
\newcommand\jcap{{JCAP}}  
\newcommand\nar{{New A ReV.}}  

\title{Detection of virial shocks in stacked Fermi-LAT clusters
}

\ShortTitle{}

\author{\speaker{Ido Reiss}\\
        Physics Department, Ben-Gurion University of the Negev, Israel\\ Physics Department, Nuclear Research Center Negev, Israel
 \\
        E-mail: \email{reissi@post.bgu.ac.il
}}

\author{Jonathan Mushkin\\
	Physics Department, Ben-Gurion University of the Negev, Israel
}

\author{Uri Keshet\\
        Physics Department, Ben-Gurion University of the Negev, Israel\\
                E-mail: \email{ukeshet@bgu.ac.il}}

\abstract{
  In the hierarchical paradigm of structure formation, galaxy clusters are the largest objects ever to virialize. They are thought to grow by accreting mass through large scale, strong virial shocks. Such a collisionless shock is expected to accelerate relativistic electrons, thus generating a spectrally flat leptonic virial ring.
However attempts to  detect virial rings have all failed, leaving the shock paradigm unconfirmed. Here we identify a virial $\gamma$-ray signal by stacking Fermi-LAT data for 112 clusters, enhancing the ring sensitivity by rescaling clusters to their virial radii and utilizing the anticipated spectrum. In addition to a central unresolved, hard signal (detected at the nominal $5.8\sigma$ confidence level), probably dominated by active galactic nuclei, we identify ($5.9\sigma$) a bright, spectrally flat $\gamma$-ray ring at the expected shock position. It corresponds to $\sim 0.6\%$ (with an uncertainty factor $\sim2$) thermal energy deposition in relativistic electrons over a Hubble time. This result validates the shock paradigm, calibrates its parameters, and indicates that the cumulative emission from such shocks significantly contributes to the diffuse extragalactic $\gamma$-ray and radio backgrounds.
}

\FullConference{7th Fermi Symposium 2017\\
		15-20 October 2017\\
		Garmisch-Partenkirchen, Germany}

\begin{document}

\subsection*{Introduction}
With a mass $M$ in excess of $10^{13}M_\odot$, galaxy clusters are located at the nodes of the cosmic web, where they accrete matter from the surrounding voids and through large-scale structure (LSS) filaments. They are thought to grow by accreting gas through strong, collisionless, virial shocks, surrounding each cluster.
These shocks form as the accreted gas abruptly slows down and heats to virial temperatures.

Strong collisionless shocks are thought, by analogy with supernova remnant (SNR) shocks, to accelerate charged particles to highly relativistic, $\gtrsim 10\TeV$ energies.
These particles, known as cosmic ray (CR) electrons (CREs) and ions (CRIs), are accelerated to a nearly flat, $E^2dN/dE\propto \const.$ spectrum (equal energy per logarithmic CR energy bin), radiating a distinctive non-thermal signature which stands out at the extreme ends of the electromagnetic spectrum, in high energy \gama-rays\cite{LoebWaxman00, TotaniKitayama00,KeshetEtAl03} and in other\cite{KeshetEtAl04, KushnirEtAl09, YamazakiLoeb15} bands.
High-energy CREs cool rapidly, on timescales much shorter than the Hubble time $H^{-1}$, by Compton-scattering cosmic microwave-background (CMB) photons.
These up-scattered photons should then produce \gama-ray emission in a thin shell around the galaxy cluster, as anticipated analytically\cite{LoebWaxman00, TotaniKitayama00, WaxmanLoeb00, KeshetEtAl03} and calibrated using cosmological simulations\cite{KeshetEtAl03, Miniati02, KeshetEtAl04}.

Once the energy accretion rate $\dot{M}T$ of the cluster has been determined, 
its \gama-ray signature depends on a single free parameter, namely the CRE acceleration efficiency $\xi_e$, defined as the fraction of downstream thermal energy deposited in CREs.
As high-energy CREs are short lived, the \gama-ray signal should reflect their spatially- and temporally-variable injection rate.
Locally, the signal thus depends on the single free parameter $\xi_e\dot{m}$, where $\dot{m}\equiv \dot{M}/(MH)$ is the dimensionless mass accretion rate and $H$ is Hubble's constant.

A direct search for a virial shock is challenging, with only the Coma cluster showing virial signals in VERITAS\cite{KeshetEtAl12_Coma}, Fermi and ROSAT\cite{KeshetReiss17} data.
A more powerful approach is to boost the virial shock signal by stacking the data of many different clusters. However, until now this method failed to indicate a robust virial shock signal.
Attempts to stack the 
Fermi-LAT data\cite{ReimerEtAl03,AckermannEtAl10, AckermannEtAl14_GammaRayLimits, HuberEtAl13, ProkhorovChurazov14, ZandanelAndo14, GriffinEtAl14} failed to find a virial signal, although they did identify emission from the centers of clusters\cite{AckermannEtAl14_GammaRayLimits,ProkhorovChurazov14} and from their large-scale environment\cite{BranchiniEtAl17}, associated with active Galactic nuclei (AGN).

\subsection*{Data preparation.}
We use the archival, $\sim8$ year, Pass-8 LAT data from the Fermi Science Support Center (FSSC)\footnote[2]{\texttt{http://fermi.gsfc.nasa.gov/ssc}}, and the Fermi Science Tools (version \texttt{v10r0p5}).
Pre-generated weekly all-sky files are used, spanning weeks $9\till422$ for a total of $414$ weeks ($7.9\yr$), with ULTRACLEANVETO class photon events.
A zenith angle cut of $90\dgr$ was applied,
according to the appropriate FSSC Data Preparation recommendations.
Good time intervals were identified using the recommended selection expression \texttt{(DATA\_QUAL==1) and (LAT\_CONGIF==1)}. Sky maps were discretized using a HEALPix scheme\cite{GorskiEtAl05} of order $N_{hp}=10$, providing a mean $\sim 0.057\dgr$ pixel separation. Event energies were logarithmically binned into $N_\epsilon=4$ energy bands in the (1--100) GeV range.
Point source contamination was minimized by masking pixels within the $90\%$ containment angle of each point source in the LAT 4-year point source catalog (3FGL)\cite{FermiPSC}.
In order to reduce the Galactic foreground, we mask $|b|<20\dgr$ latitudes, near the 
bright Galactic plane.

\begin{figure}[h]
	\centerline{
		\includegraphics[trim={2.7cm 1.2cm 1.5cm 1cm}, clip, width=14.4cm]{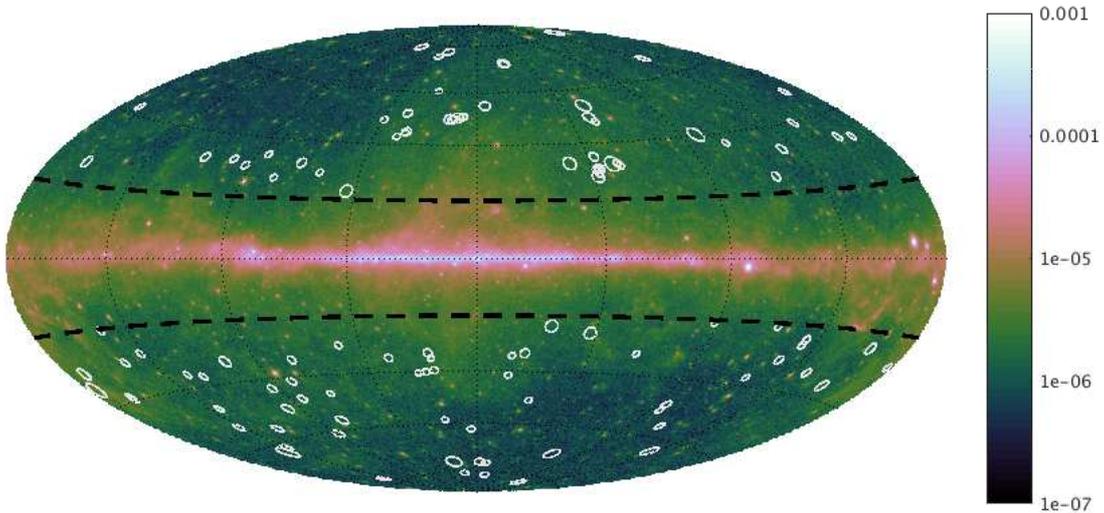}
	}
	\caption{\label{fig:cluster mapA}
		Fermi-LAT photon flux (in units of $\se^{-1} \cm^{-2} \sr^{-1}$) sky map in the $(1\till500)\GeV$ energy range, shown in a Hammer-Aitoff projection with Galactic coordinates. The locations (white circles of radius $5\myR$) of the 112 clusters used in the analysis are superimposed.	
	}
\end{figure}

We stack the LAT data around 112 clusters (see Fig.~\ref{fig:cluster mapA}) selected from the Meta-Catalog of X-ray Clusters\cite{PiffarettiEtAl11} according to the following criteria: \myNi a mass $M_{500}>10^{13}M_\odot$ enclosed within $\myR$; \myNii an angular radius $0.2\dgr<\theta_{500}<0.5\dgr$, chosen to avoid small angles below the high-energy LAT PSF, and extended clusters where the foreground estimation is complicated; \myNiii a sufficient distance from the Galactic plane, with latitude $|b|>20\dgr$; and \myNiv a distance of at least $1.8\dgr$ (the $90\%$ containment angle at $1\GeV$) from any 3FGL point source.

\vspace{-0.3cm}
\subsection*{Direct significance estimation.}
Cluster virial radii span a wide range of spatial ($r_v$) and angular ($\theta_v$) scales. Hence, unlike previous studies, we select and stack the data on the normalized angles $\mytautheta\equiv\theta/\theta_{500}$. 

The foreground, after point sources and the Galactic plane were masked, varies mainly on scales much larger than the anticipated extent of the cluster signal. Therefore, this remaining foreground can be accurately approximated using a polynomial fit on large scales. For each cluster, we thus consider an extended, $0<\mytautheta<\mytautheta_{max}\equiv 15$ disk region around its center, and fit the corresponding LAT data by an order $N_f=4$ polynomial in the angular coordinates $\mytautheta_x$ and $\mytautheta_y$.
This is done separately for each of the four energy bands.

\begin{figure}[h!]
	\centerline{
		\includegraphics[width=14.0cm]{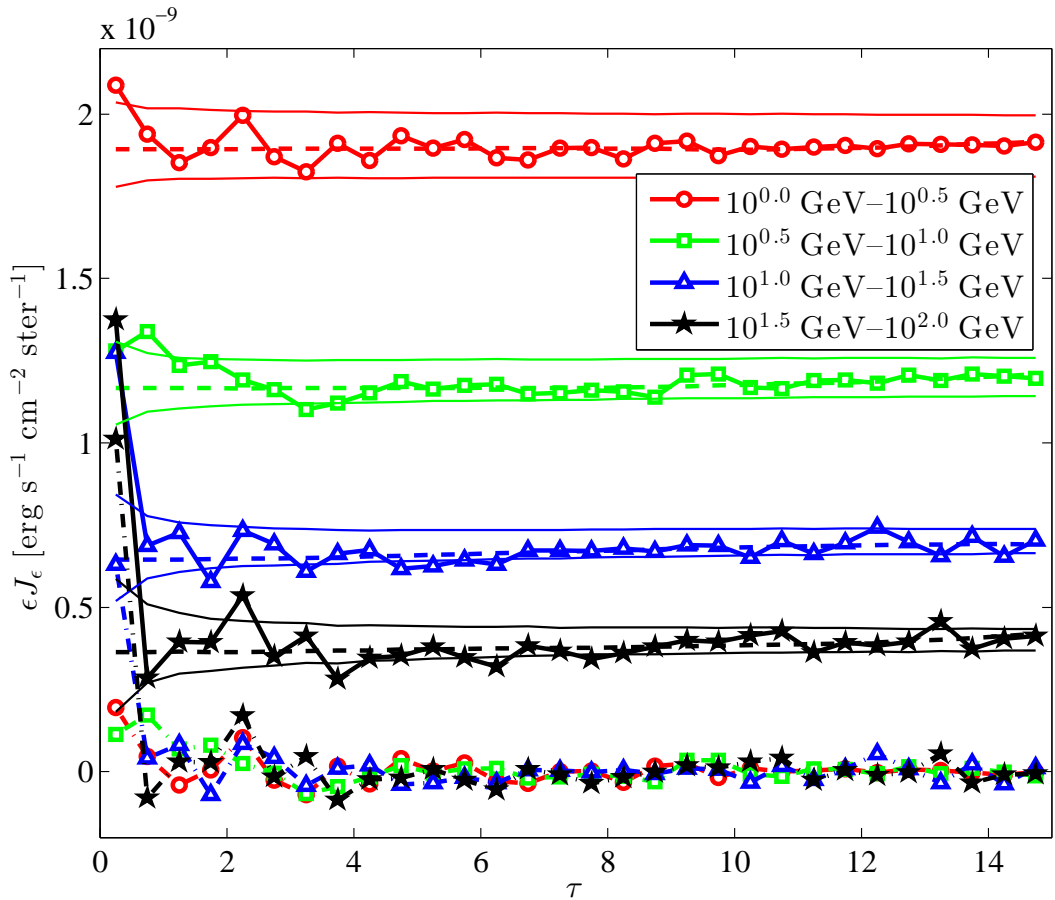}
	}
	\caption{\label{fig:flux} Scaled and radially binned energy flux, stacked over the full cluster sample, as a function of the normalized radius $\tau$, shown in each of the four energy bands (symbols given by the legend, with solid lines). Also shown are the estimated foreground (dashed curves) and the excess emission (lower symbols, with dash-dotted lines).
	}
\end{figure}

For each cluster $c$, each photon energy band $\epsilon$, and each radial bin centered on $\mytautheta$ with width $\Delta \mytautheta=0.5$, we define the excess emission $\Delta n\equiv n-f$, where $n$ is the number of detected photons, and $f$ is the number of estimated foreground photons.
The resulting stacked flux, foreground flux, and excess emission are shown in Fig. \ref{fig:flux}. The significance of the excess emission can be estimated, assuming Poisson statistics with $f\gg1$, as
\begin{equation} 
	\label{eq:SingleBinSignificance}
	\nu_{\sigma,c}(\epsilon,\mytautheta) = \frac{\Delta n_c}{\sqrt{f_c}} \fin
\end{equation}

Next, we stack the data over the clusters in the sample.
To examine the robustness of our analysis and possible biases by a large number of photons arriving from a few high-foreground or bright clusters, or from a high significance signal arriving from a few low foreground clusters, we define two different methods to compute the significance of the signal stacked over clusters.

The first, more standard method is photon co-addition.
Here, at a given radial bin and energy band, we separately sum the excess photon count and the foreground photon count over the $N_c$ clusters.
The stacked significance is evaluated as the ratio between the stacked excess and the square root of the stacked foreground,
\begin{equation}
	\nu_\sigma^{(ph)}(\epsilon,\mytautheta) = 
	\frac
	{\sum_{c=1}^{N_c} \Delta n_c}
	{\sqrt{\sum_{c=1}^{N_c} f_c}} \fin
\end{equation}
The second method is cluster co-adding.
Here, at a given radial bin and energy band, we co-add the significance $\nu_{\sigma,c}$ of Eq.~(\ref{eq:SingleBinSignificance}) over the $N_c^*(\epsilon,\mytautheta)$ clusters for which it is defined (\ie where $f_c>0$),
\begin{equation} \label{eq:clusterCoAddition}
	\nu_\sigma^{(cl)}(\epsilon,\mytautheta) = 
	\frac{\sum_{c=1}^{N_c^*} \nu_{\sigma,c}}
	{\sqrt{N_c^*}} \fin
\end{equation}
The two methods qualitatively agree with each other, although they do differ in a handful of bins by up to $\sim 1\sigma$. The difference between the two methods gauges the stacking systematics. 

\begin{figure}[h]
	\centerline{
		\includegraphics[width=14.0cm]{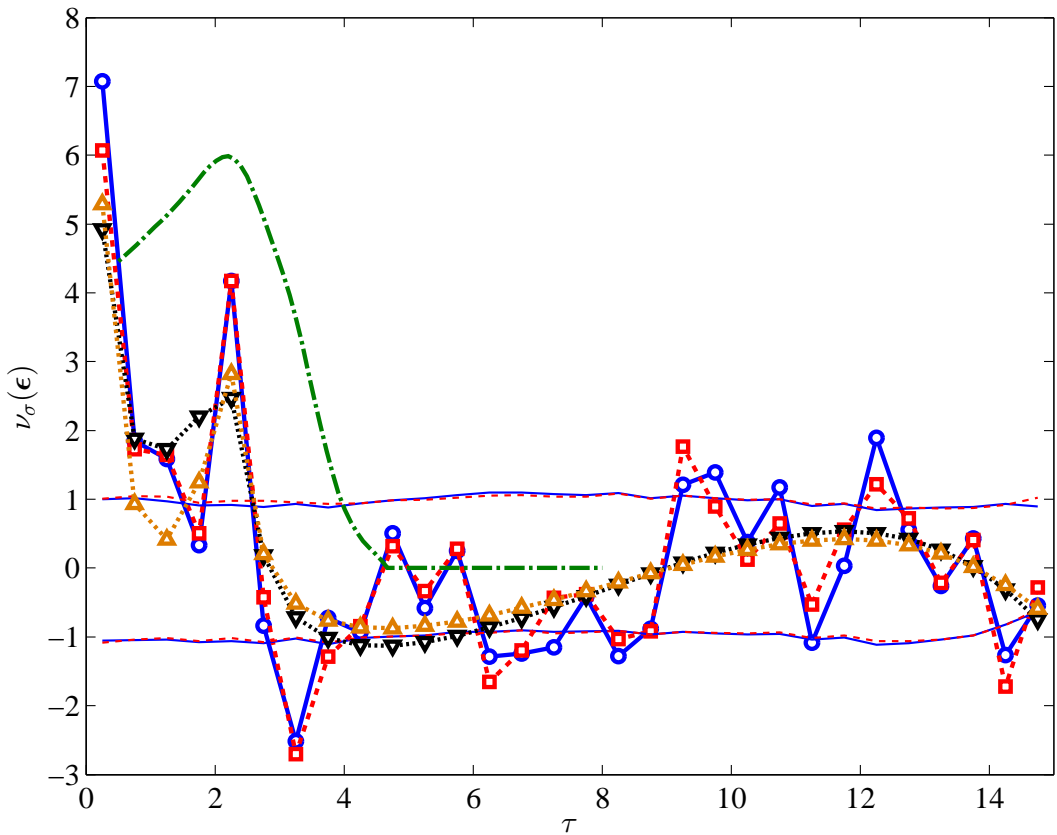}
	}
	\caption{\label{fig:combinedA}
		Significance $\nu_\sigma(\bm{\epsilon})$ estimation as a function of the scaled angular radius $\tau$, stacked both by photon co-addition (blue circles with solid lines) and by per-cluster signficance co-addition (red rectangles with dashed lines). The $1\sigma$ confidence levels of the mock catalog distributions are shown (thin lines) for photon co-addition (solid blue curve) and for cluster co-addition (dashed red). Also shown are the simulated signals for the best fit models combining AGN with a full virial shock (black down-triangles with a dotted line) and with a planar shock (orange up-triangles with a dotted line). The TS-equivalent signficance values of the full leptonic ring are also shown (green dash-dotted line).
	}
\end{figure}

Next, we co-add the $N_\epsilon=4$ logarithmic energy bands with equal weights,
\begin{equation} \label{eq:BandCoAdd}
	\nu_\sigma(\bm{\epsilon},\mytautheta) = 
	\frac{\sum_{\epsilon=1}^{N_\epsilon}\nu_{\sigma}(\epsilon,\mytautheta)}
	{\sqrt{N_\epsilon}} \fin
\end{equation}
The resulting significance of the excess emission (Fig. \ref{fig:combinedA}) shows two spatially separated components: a central component and a peripheral, ring-like component.
The two components, each arising from the cumulative contribution of many clusters, are found in a wide range of cluster masses. 

In order to validate the foreground-based significance estimation and to examine possible systematic biases, we prepare and analyze a large number ($N_{mock}=2000$) of control (mock) cluster catalogs. In each mock catalog we use the exact same cluster masses and angular radii as those in the true sample, but place them in random locations on the sky, assuring that the mock clusters satisfy the same cut criteria of the true cluster sample.

The $1\sigma$ band of the mock clusters is shown in Fig. \ref{fig:combinedA}. The mean $\langle \nu_\sigma\rangle$ of the mock catalogs deviates from zero by no more than $0.1$, revealing no large systematic bias. The variance $\mbox{Var}(\nu_\sigma)$ of the mock catalogs deviates appreciably from unity only beyond $\mytautheta\simeq 12$, indicating that out to this large radius 
, our significance estimates are reliable. 

\subsection*{Parameter and TS-based significance estimation.}

In order to measure the model parameters and their uncertainty, and to accurately determine the significance of the signals, one must take into account the PSF corrections, the signal and foreground photon statistics, and the correlations that are induced by cuts in the map, by masked pixels, and by our methods of stacking.
We do so primarily using a control sample, Monte Carlo simulating the LAT data that would arise from the clusters of a mock catalog, for a given choice of model parameters (based on cluster $\beta$-models; see [\!\!\citenum{ReissEtAl17}]).
The resulting mock photon counts are then injected into the real LAT data, and the result is analyzed with the same pipeline used to study the real clusters.
We repeat this for $N_{mock}=10$ catalogs, and for a large set of parameter values.
Each mock cluster corresponds to a real cluster in our sample, and is assigned with the same parameters but with a random location in the permitted region of the sky. A maximal likelihood (minimal $\chi^2$) analysis is used to calibrate the model and estimate the uncertainties in the parameters, and the significance is estimated using the test statistics\cite{MattoxEtAl96_TS} TS.

\subsection*{Results.}
The central emission is unresolved, confined to the innermost $0.5\theta_{500}$, where it presents at an energy co-added significance $6\sigma\till7\sigma$.
It is morphologically consistent with a point source located at the center of the cluster.
Fitting a point-source model indicates a hard, $\mySagn=-1.61^{+0.24}_{-0.18}$ photon spectral index, consistent with AGN.
The signal can be crudely interpreted as one out of every four clusters in our sample harboring a point source of luminosity $\myLAGN\sim 6\times10^{41} \erg \se^{-1}$ in the emitted $(1\till100)\GeV$ band.
These conclusions support and extend previous claims for a faint population of \gama-ray AGN\cite{ProkhorovChurazov14, BranchiniEtAl17}.

The peripheral, ring-like signal peaks at $(2.0\till2.5)\theta_{500}$, where it presents at a significance of $4.2\sigma$. 
This signal matches the expected signature of \gama-ray rings arising from inverse-Compton scattering of CMB photons by virial-shock accelerated CREs.
Fitting a virial ring indicates CREs injected at a mean scaled shock radius $\mytaur_v\equiv r_v/R_{500}=2.4\pm0.1$, at a rate $\xi_e\dot{m}=(0.65\pm0.11)\%$.
The signal is consistent with a flat, $s_v=-2$ photon spectral index, and changes little when modelling, masking, or removing the central sources; a ring-only model gives $s_v=-2.10^{+0.20}_{-0.16}$.
The calibrated model is consistent with the data in all four energy bands and in the four equal logarithmic mass bins used in the fit, and is consistent with the Coma signal  \cite{KeshetReiss17} (which was excluded from our sample) and previous upper limits. 

Our nominal significance estimates, based on the test statistics\cite{MattoxEtAl96_TS} TS in our binned stacking method, are $5.9\sigma$ for the virial ring, and $5.8\sigma$ for the AGN.
We carry out a suite of convergence and sensitivity tests, indicating that our results are robust to variations in the preparation of the LAT data (point source and Galactic plane masking) and of the cluster sample (cuts on mass, angular radius, and proximity to point sources and to the Galactic plane), in the photon analysis methods (discretization, foreground modelling), in the cluster stacking methods (photon vs. per cluster significance co-addition, different mass bin co-additions, radial bin size), and in our energy co-addition method (number of energy bins).

To test if the ring signal is narrower (in $\tau$) than the model, as may  apparently (but not significantly, according to the $\chi^2$ values) seem from Fig. \ref{fig:combinedA}, we test if the stacking may have preferentially picked up shocks with brighter emission in the plane of the sky (as inferred in Coma \cite{KeshetReiss17}). 
This model yields a signature (shown in the figure as down triangles) of width comparable to the signal, but of nominal significance ($4.2\sigma$) lower than that of the spherical shock model, and is therefore currently disfavored.

\subsection*{Conclusions}

We presented a robust, high significance LAT detection of virial rings around stacked galaxy clusters, consistent with the signals inferred around Coma.
This confirms the paradigm of LSS accretion through virial shocks.
The shocks are not highly non-spherical, otherwise the signal would have been smeared by the stacking.
They are consistent with a nearly fixed enclosed over-density $\delta$, as accordingly rescaling their radius has facilitated the detection of the stacked signal;
furthermore, the shock location closely matches that expected from simple spherical collapse models \cite{EkeEtAl96}, and from simulated $\Lambda$CDM clusters\cite{KeshetEtAl03, SchaalSpringel15}.
Our results positively test the theory of CRE acceleration, generalizing it to scales much larger than accessible ever before.

Adopting our nominal CRE injection rate as typical of all clusters, we obtain\cite{KeshetEtAl12_Coma} a diffuse \gama-ray component $\epsilon^2 dJ/d\epsilon\simeq 0.1(\xi_{e}\dot{m}/1\%)\keV\se^{-1}\cm^{-2}\sr^{-1}$, 
contributing a significant fraction of the extragalactic\cite{KeshetEtAl04_EGRB} \gama-ray background.
In the radio, we find a $\nu I_\nu \sim 10^{-11}(\xi_{e}\dot{m}/1\%) (\xi_B/1\%)\erg\se^{-1}\cm^{-2}\sr^{-1}$ synchrotron signal, observable through $\delta T_l\simeq 0.4 (\nu/\mbox{GHz})^{-3}\K$ fluctuations at multipoles $400\lesssim l \lesssim 2000$ with present interferometers such as LOFAR and EVLA\cite{KeshetEtAl04, KeshetEtAl04_SKA}.

\end{document}